\documentstyle[12pt]{article}

\title{\bf On geometric phase from pure projections}

\vspace{60mm}

\author{\bf Rajendra~Bhandari}

\date{ }

\begin{document}

\maketitle
\vspace{40mm}
\begin{center}
\begin{tabular}{ll}
            & Raman Research Institute, \\
            & Bangalore 560 080, India. \\
            & email: bhandari@rri.ernet.in\\
\end{tabular}
\end{center}
\vspace{35mm}
-----------------------------------------------------------------------\\
submitted to Journal of Modern Optics.\\
MS version of 4 February 1998
\newpage
\begin{center}
{\bf Abstract}
\end{center}
\vspace{1mm}

The relation between 
three recent experiments aimed at demonstrating 
a strictly geometric phase using pure projections
is clarified in the light of a recent claim to such
a measurement. The possibility of observing  $\pm 2\pi$
phase jumps in an optical interference experiment is
pointed out.

\newpage
\section{Introduction}

Recently, Hariharan et.al. \cite{hrss} have  reported an
experiment (HRSS Expt.) in which they extend the work of 
Berry and Klein \cite{berklein} (BK Expt.) to measure a 
pure geometric phase
arising from state projections, as formulated by
Pancharatnam \cite{panch1}. These experiments are claimed \cite{hrss}
to be  intrinsically free from all sources of
dynamical phase.

In this paper I show that (i) there are difficulties in
regarding the  HRSS Expt. as a measurement
of a pure geometric phase since it is not intrinsically free from
dynamical phase as claimed, (ii) an experiment earlier
reported in literature \cite{iwbs} (the IWBS Expt.) 
did measure a pure geometric phase arising from state 
projections. In addition to measuring a linearly increasing
geometric phase, it measured in detail the nonlinear behaviour
and  jump of the geometric phase through $\pm \pi$
near a singularity and explicitly verified the existence
of the  singularity associated with the
geometric phase, (iii) the BK Expt. and the
HRSS Expt. contain similar singularities which could be seen 
if the techniques used and the available sensitivities   
allowed a more complete sampling 
of the parameter spaces in these experiments
and (iv) in the HRSS expt. one could, in principle, 
see phase jumps equal to $\pm 2\pi$, a somewhat surprising result.

\section{Historical remarks}

Following Berry's discovery of the adiabatic phase \cite{berry1},
there was a widespread revival of interest in Pancharatnam's
work on polarization of light \cite{panchcoll}, in particular,
in the form of experimental demonstrations of the geometric
phase by interferometry, the first one being an experiment by
Bhandari and Samuel \cite{rbjs}. This was followed by a number
of experiments by other groups \cite{nonint,sks,mandel,tompkin,nonclassical}. 
A common feature of all these experiments was
that these involved circuits on the Poincar\'{e} sphere consisting
of geodesic curves, caused by unitary transformations.
For  a unitary transformation which takes the state along a geodesic,
the dynamical phase is zero \cite{berry2}, 
as in case of circuits caused by projections, considered by 
Pancharatnam  \cite{panch1}. The geodesic
nature of the circuits is ensured in these experiments by
appropriate choice of the polarization elements like retarders
and their orientation.  Departures from a geodesic circuit
would arise in these experiments only from errors in the retardation 
and orientations. The other common feature of these experiments
was that these measured a  geometric phase increasing linearly
as a function of an experimental parameter, typically rotation
angle of an optical element. A more recent measurement of such
a linear phase using neutron interferometry has been reported by
Wagh et.al.\cite{wagh}.

Hariharan and Roy  reported a new version
of the above experiments  in which they use a
Sagnac interferometer configuration \cite{hari}. Since the 
two interfering
beams in this experiment travel the same physical path in
space, all U(1) phase errors (errors in the polarization-independent 
part of the phase)
are eliminated in this experiment. However since the beams
still encounter a beam splitter and mirrors this experiment
is not free from SU(2) phase errors caused by unknown
polarization changes brought about by these optical elements.
Moreover this experiment does not measure fringe shifts. It
measures the modulation of 
intensity of the superposed beams as a function of
orientation of a halfwave plate which is interpreted as a
linearly increasing {\it difference} of the geometric 
phase shifts acquired by the clockwise and the counterclockwise
beams.

Following this, the present author reported an experiment
\cite{iwbs} (the IWBS Expt.) that directly measures phase shifts 
in which the two interfering beams, 
with orthogonal polarizations, are not split , hence they
do not encounter any beam splitter or mirrors 
\footnote{The label IWBS stands for
"Interferometry Without Beam Splitters"}. This eliminates 
the SU(2) errors as well. Yet the two beams 
acquire a geometric phase difference. This idea was incorporated by
Hariharan and Roy in their new technique for phase stepping
interferometry \cite{stepping,haritalk}.

\section{The IWBS Experiment}

This experiment \cite{iwbs} is performed with a Hewlett Packard laser 
interferometer that uses heterodyne techniques to  measure 
and record phase shifts 
between two light waves at He-Ne laser frequency with a
sensitivity of $\lambda
/20$, alongwith its sign. It uses 
a novel interferometer configuration in which the two
interfering beams with orthogonal polarizations, represented
by the points $A$ and  $\tilde{A}$ on the Poincar\'{e} sphere
(figure 1), travel along
the same path in space, pass through a quarterwave plate
so oriented as to take the two beams to a chosen pair of orthogonal
polarization states represented by the points $P$ and  $\tilde{P}$
and then pass through a linear polarizer which passes the
state represented by N (figure 1). The phase difference
between the beams is continuously recorded as a function
of rotation of the polarizer resulting in rotation of the
point N on the sphere. The experiment is performed for
several orientations of the quarterwave plate P which
determine the polar angles $\theta$ and $180^\circ-\theta$
of the points $P$ and $\tilde{P}$. The phase shift 
measured in this experiment as the polarizer is rotated from
Q to N is given by  the phase of the
complex number $<P\mid Q><Q\mid \tilde{P}><\tilde{P}\mid N><N\mid P>$,
where $\mid P>,\mid N>,\mid \tilde{P}>,\mid Q>$ are respectively
the two-component spinors representing the states $P,N,\tilde{P},
Q$ and is equal to half the solid angle subtended by the shaded area
bounded by the two geodesics $PNF\tilde{P}$ and $\tilde{P}LQP$.
It is thus a pure geometric phase arising from four projections.

Figure 2 reproduces from ref.\cite{iwbs} the expected phase 
shift (solid line) and
the measured values (dots) as a function of the orientation
of the linear polarizer for several different values of
$\theta$. For $\theta = 0^\circ$ and $\theta = 180^\circ$,
the phase shift is linear and corresponds to the 
situation studied in the HRSS expt. For other values of
$\theta$, the phase shift is nonlinear and for values 
close to $\theta = 90^\circ$, the curves are highly
nonlinear near critical orientations of the polarizer,
switching sign at $\theta = 90^\circ$. The singular
behaviour occurs when the state $\mid N>$ becomes
orthogonal to the state $\mid P>$ or $\mid \tilde{P}>$
and one of the scalar products in the above product becomes
zero. The jump in the phase near the singularity in these
experiments is equal to $\pm \pi$.
Such jumps were first
predicted in ref. \cite{jumps} and their 
relevance to the phase evolution of quantum symstems 
pointed out in refs. \cite{jumps,review}.

Another independent observation of the nonlinearity of
the Pancharatnam phase was reported by 
Schmitzer et. al. \cite{schmitzer}.
This experiment, however, does not detect the switch in the
sign of the phase and the singularity. 

\section{The HRSS Experiment}

In this experiment \cite{hrss}, the authors use a Sagnac
interferometer configuration with a nonpolarizing beam 
splitter and three mirrors and a rotatable linear analyser 
LA placed between two of the mirrors. A He-Ne laser 
beam rendered linearly polarized by means of a linear
polarizer LP (fig. 1 of ref. \cite{hrss}) is then converted 
into left circular by
means of a quarterwave plate before incidence on the
beam splitter where it is divided into a clockwise and
a counterclockwise beam. The beams then pass through a
linear polarizer LA in the opposite directions and 
recombine at the beam splitter. The two beams then pass 
through a common right circular analyzer and their combined
intensity sensed with a photodiode. The variation of
this intensity at a given point in space is monitored
as a function of the orientation of LA, i.e. $\phi/2$
(or $\theta$ in the notation of ref.1). 
From an observed sinusoidal variation of the intensity, the
authors infer a {\it change} in geometric phase 
linear in $\phi$.

Such an interpretation of the observed intensity curve
has several problems. First of all it is not equivalent to
measurement of a phase shift as a function of $\phi$.
This would be so if for example a spatial shift of interference fringes
on a screen were measured directly. Alternatively, phase shifts
could be measured by heterodyne interferometry where these could be
looked upon as fringe shifts in the time domain. Secondly, it is not
true that the beams do not undergo any SU(2) transformation
during the passage through the interferometer. Every reflection
from a mirror or a beam splitter introduces an SU(2)
transformation on the polarization state of the beam.
For reflection off an ideal metal mirror, this could be
taken to be equivalent to that due to a halfwave plate \cite{decomp}.
In practice the reflections are not ideal.
Beam splitters in particular are notorious for introducing
undesirable polarization changes. As the linear analyzer
LA is rotated, a variable polarization is incident on the
beam splitter and an SU(2) transformation varying with
$\phi$ would be introduced. The evolution due to pure
projections is therefore accompanied by SU(2) evolution.
This introduces a dyamical phase error which is a function of
the orientation of the polarizer, hence significant.
Since fringe shifts are not measured, the phase errors
due to these effects cannot be easily estimated. 

Finally, we show that if the experiment were performed in
the fringe-shift  mode and adequate sensitivity
available, one could see a singularity in this experiment
too. Let the initial polarization state in the experiment
be chosen to be an arbitrary elliptically polarized state
shown as the point N in figure 3. This could be achieved
simply by orienting the first polarizer LP 
(fig. 1 of ref.\cite{hrss}) suitably,
without disturbing the quarterwave plate. Let $\theta$
be the polar angle of the point N. The circuits traced
by the two beams are NPL and NQL, where NP, PL, NQ and QL
are geodesic curves and the interferometer
records the phase equal to half the solid angle of the
circuit NPLQN, which is equal to the phase of the complex
number  $c=<N\mid P><P\mid L><L\mid Q><Q\mid N>$.
This changes nonlinearly in general as the polarizer LA is 
rotated.

Figure 4 shows, for different values of $\theta$, the
computed phase shifts expected in such an experiment as 
a function of the orientation
of the analyzer LA (fig. 1 of ref. \cite{hrss}).
The curve for $\theta =0^\circ$, a straight line, corresponds to 
the experiment actually reported. The interesting feature is the
$\pm 2\pi$ jump  near $\theta =90^\circ$ and 
$\phi =90^\circ$ (see curves C and D).
This is a new result and in fact contradicts the conclusion
reported in ref.\cite{review}, namely in polarization
interference experiments one sees only $\pm \pi$ phase jumps.
The earlier conclusion holds in the context of experiments where
only one of the scalar products forming the product $c$ goes to 
zero at a time. In the present experiment, at the singular point 
in the parameter space, two scalar products in the product go
to zero simultaneously. This can double the strength of the
singularity.

\section{The BK Experiment}

In one of the experiments reported in ref. \cite{berklein},
Berry and Klein consider propagation of polarized light
through a sequence of N polarizers that
transmit states $\mid \bf{r_1}>$, $\mid \bf{r_2}>$, ...,
$\mid \bf{r_{N}}>$, represented respectively by points
$\bf{r_1}$, $\bf{r_2}$, ... , $\bf{r_{N}}$ on the
Poincar\'{e} sphere which are distributed uniformly on a 
small circle with polar angle $\theta$. Under such a
propagation, which is equivalent to N projections,
the beam acquires a pure geometric phase which, for large N, 
is equal
to the solid angle subtended by the small circle at the
centre of the sphere. The actual experiment is done 
for N=4 and $\theta =90^\circ$, i.e. with 4 linear polarizers.
The authors observe a fringe shift of magnitude $\pi$,
as expected.

Let us note that the passage of a light beam through
a polarizer is also accompanied with a large dynamical
phase change of the U(1) type due to the refractive
index of the polarizer material being different from 1. 
In order to see only the phase shift due to the projections,
the above dynamical phase must be reliably compensated 
with residual error small compared to the geometric 
phase being observed. What is observed in the present
experiment is, therefore, not the phase difference but 
the {\it change} in 
phase difference between the two beams of the interferometer
as the polarizers are rotated from a configuration 
when their easy axes are aligned to the final
configuration. Presumably it is ensured beforehand
that a rotation of the polarizers does not cause a
change in the U(1) dynamical phase. What one observes,
therefore, is a {\it change} in geometric phase as the
circuit evolves with the rotation of the successive
plates.

The point we wish to make is that if the experiment
were performed with general polarizers corresponding
to arbitrary values of $\theta$, it is possible in
principle to keep track of the position of the fringes
continuously as the polarizers are rotated one by
one. The track of the polarization state at an
intermediate stage when three of the polarizers have
been rotated to their final configuration is 
schematically shown in figure 5 for
two values of $\theta$, one in each hemisphere.
Assuming the  phase shift to be zero when all
polarizers are aligned with the first one, the phase
shift at the intermediate stage is given by half the
solid angle subtended by one of the shaded areas
in figure 5. Note that the circuits in the upper and
the lower hemisphere are traversed in the opposite sense,
implying a phase shift of the opposite sign.
The computed values of the phase shift as a function of 
the orientation angle $\phi/2$ of the last polarizer rotated
(i.e. the azimuth $\phi$ of the point D or D')
for several values of $\theta$ are shown in figure 6,
where N = 360, a large number.
The $\pm \pi$ phase jump near  $\theta =90^\circ$ and
$\phi =180^\circ$ (see curves C and D) is thus in principle observable.
In the context of unitary transformations, an earlier
experiment \cite{4pism} aimed at demonstrating the
$4\pi$ spinor symmetry principle using polarization states,
deals with similar circuits.
The switch in the sign of the phase change in going
from the upper to the lower hemisphere has been observed
in this experiment.
Let us note that  the standard result is that the
geometric phase for a complete circuit of the above type
is equal to $\pi(1-cos\theta)$, irrespective of the value
of $\theta$. This does not contain such a jump.

\section{Summary}

Our analysis of three experiments aimed at measuring
a pure geometric phase as formulated by Pancharatnam
reveals that in every case a singularity in the
parameter space of the experiment where the phase
becomes undefined is an essential feature of the
problem.  While in the IWBS expt. the singularity
has been demonstrated in detail, in the HRSS and the
BK expts.
these are potentially measurable. In the context of
unitary transformations, such singularities have
been seen in earlier polarization experiments
\cite{dirac1,dirac2}. While all the singularities
observed so far have strength $\pm 1$ yielding a total
phase change around a circuit equal to $\pm 2\pi$, the 
singularity in the HRSS expt. has strength $\pm 2$ and 
would be interesting to observe.

\section{Note added :}

One of the referees has expressed the view (probably
based on the conclusions of ref. \cite{senarmont}) that the
principle of the IWBS experiment is the same as that
of the Senarmont compensator used in interferometers
in the sixties and the seventies. We disagree. The
Senarmont compensator falls into the class of 
polarimetric devices that sense the polarization state
of a beam with the help of retarders and polarizers.
The IWBS experiment is an interferometric technique
that senses the phase shift between two beams. The
two important constraints of equal amplitudes
in the two polarizations and a special orientation of
the quarterwave plate that takes the two linear polarizations
to the circularly polarized states in the Senarmont compensator
are not present in the IWBS experiment. It is also not
clear how nonlinear phases measured in the IWBS 
experiment would be measured with a Senarmont compensator.

\newpage

\section*{}{\Large {\bf Figure Captions}}

\addcontentsline{toc}{section}{Figure Captions}

{\bf Figure 1}:
The QWP oriented at an angle $-45^\circ+\theta/2$
 to the x-axis in real space rotates the initial states 
 $\mid A>$ and $\mid {\tilde A}>$ to the states $\mid P>$
 and $\mid {\tilde P}>$ about the point Q which represents
 the orientation of its fast axis. The linear polarizer
 brings the states $\mid P>$ and $\mid {\tilde P}>$
 along the shorter geodesics to the state $\mid N>$
 which rotates along the equator of the Poincar\'{e} sphere
 through $2\tau$ as the polarizer is rotated through
 $\tau$. The interferometer measures a phase equal to 
 half the solid angle subtended by the shaded area
 $PNF{\tilde P}LQP$, bounded by the two geodesics.

{\bf Figure 2}:
The solid line in the curves shown in (a)-(h) represents
the computed phase change as a function of $2\tau$ for a
few chosen values of the orientation of the QWP which is
equal to $-45^\circ+\theta/2$. The dots represent the 
measured values. 
The state $\mid N>$ rotates on the  sphere through
an angle $2\tau$ when the polarizer rotates through $\tau$.

{\bf Figure 3}:

The two beams in the interferometer follow the tracks
shown as NPL and NQL on the Poincar\'{e} sphere. The
measured phase difference is half the solid angle
of the area NPLQN which changes nonlinearly as the
linear polarizer LA rotates. The singularity occurs
when the initial state N is linear and coincides with
A and the angle $\phi$ equals $180^\circ$. The initial
state N is varied by changing the orientation of the
first polarizer in front of the laser.

{\bf Figure 4}:
The computed phase change as a function of the orientation
$\phi/2$ of the linear analyzer for various values of the
polar angle $\theta$ of the initial state N. Note the
$\pm 2\pi$ phase jump near the value $\theta = 90^\circ$
and $\phi/2=90^\circ$.

\newpage

{\bf Figure 5}:

The polarization state follows the track ABCDFA (A'B'C'D'F'A') 
on the Poincar\'{e} sphere at an intermediate stage when the first 
polarizer corresponds to the state A (A') and three of the 
polarizers
have been rotated to their final orientations corresponding
to the points B (B'), C (C') and D (D').

{\bf Figure 6}:

The computed phase change as a function of the orientation
$\phi/2$ of the last rotated polarizer for various values of
the polar angle $\theta$ of the circuit.  Note the geometric
phase for the full circuit flips from $+\pi$ to $-\pi$ near
$\theta=90^\circ$.
 
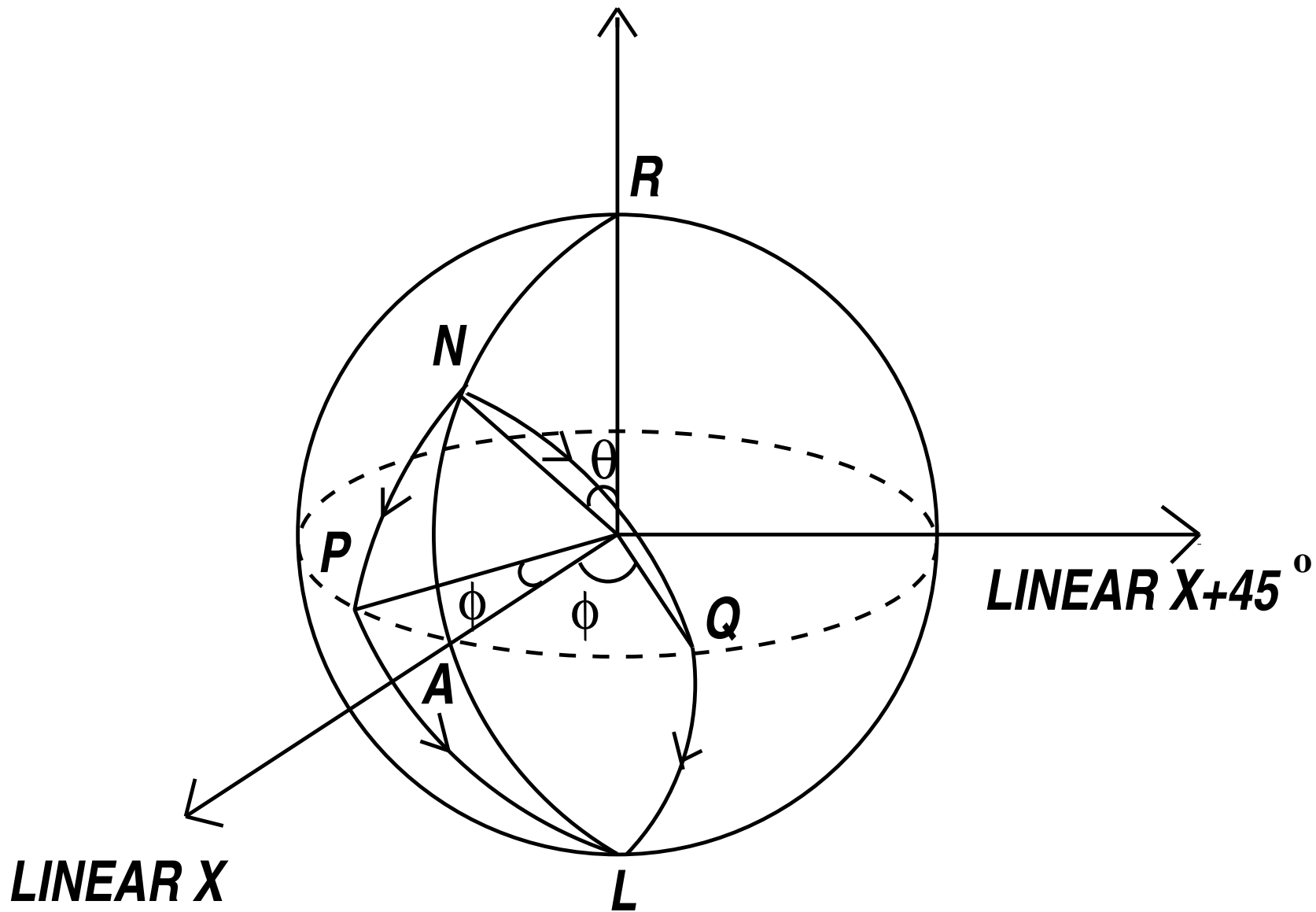
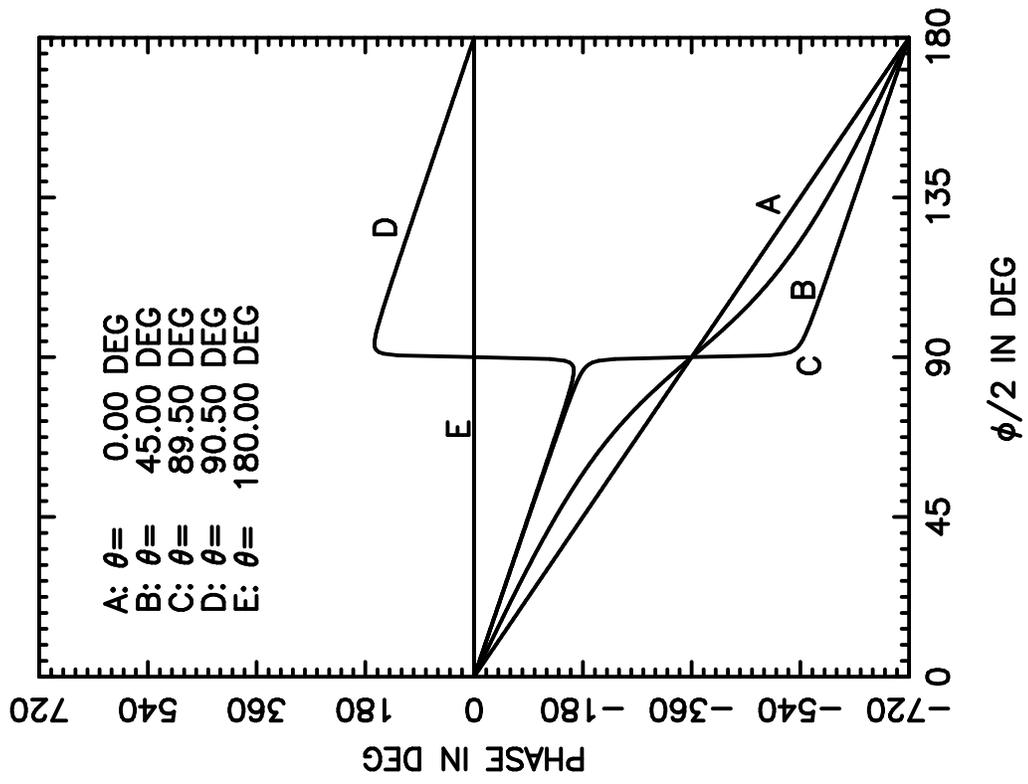
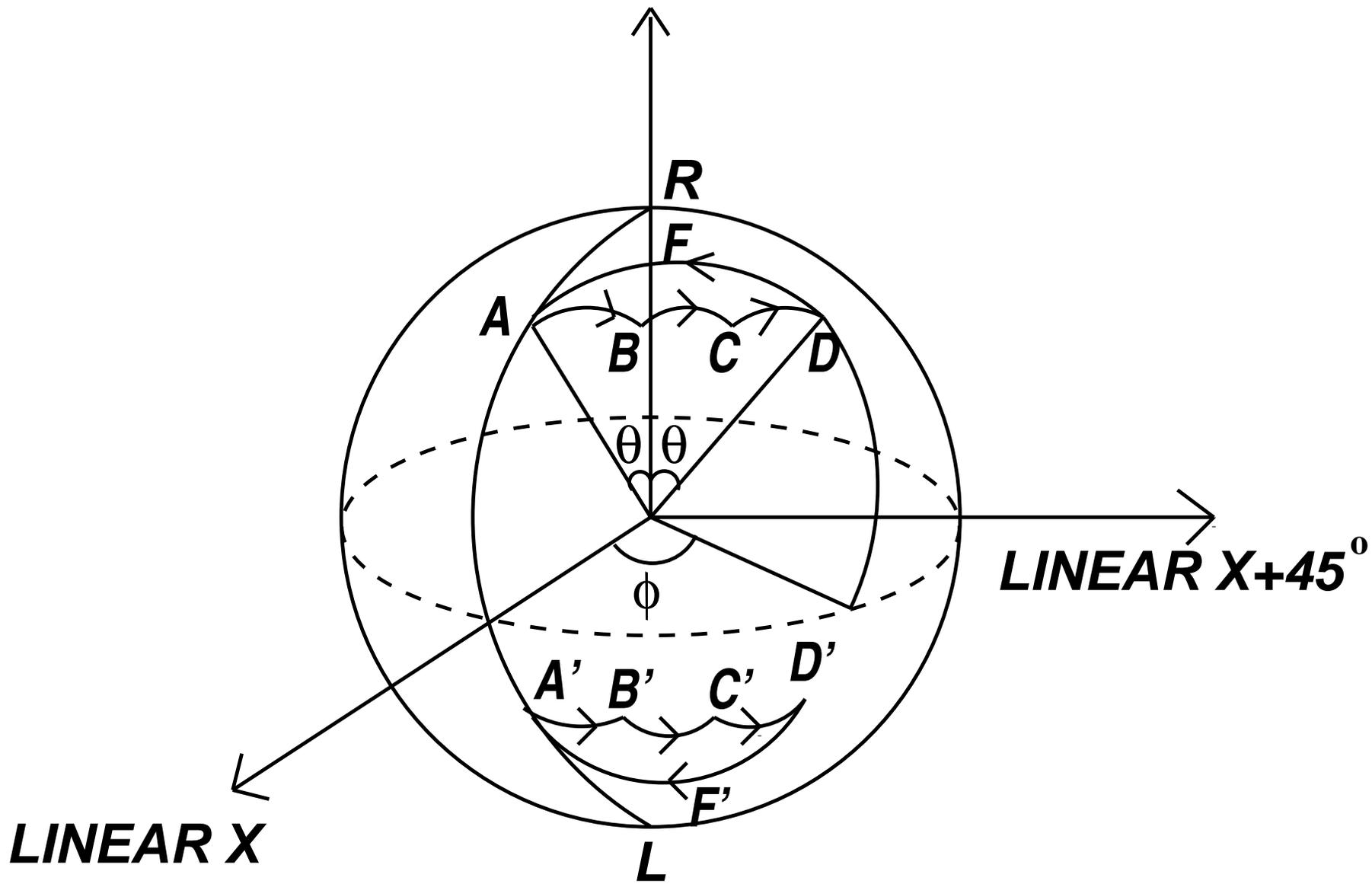
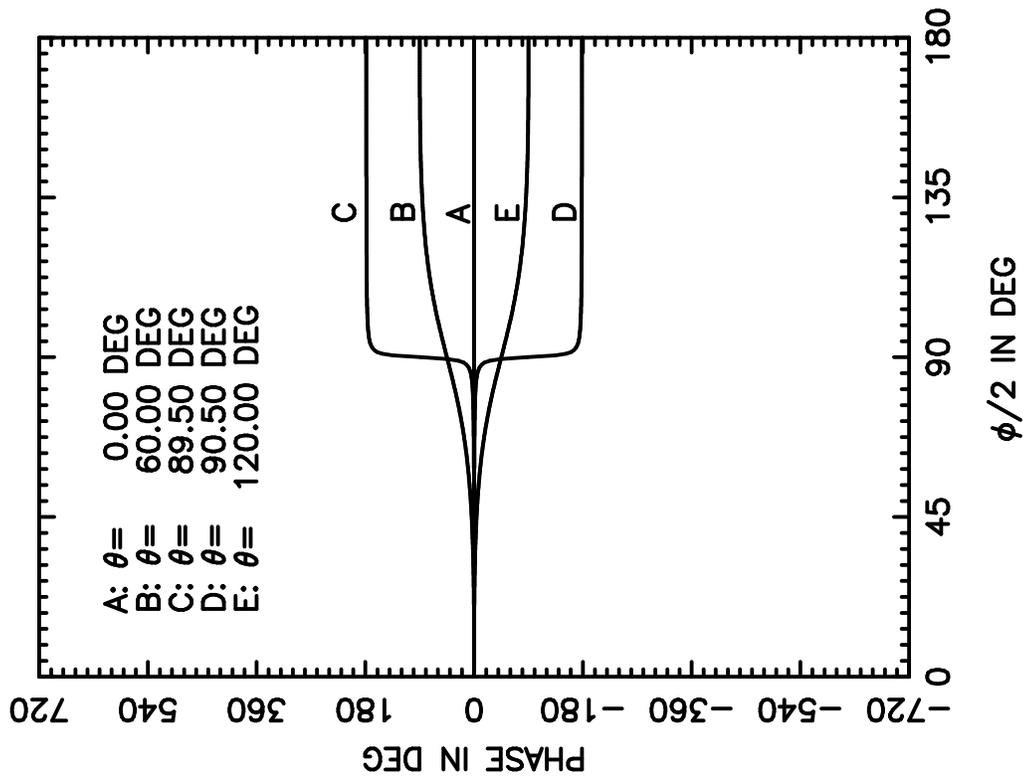
\end{document}